\documentclass[aps,pre,twocolumn,groupedaddress,superscriptaddress,showpacs]{revtex4-1}

\usepackage{graphicx}
\usepackage{color}
\usepackage{amsmath}
\usepackage{amsfonts}
\usepackage{amssymb}
\usepackage{dcolumn}

\begin{document}
  \title{Gaussian model of explosive percolation in three and higher dimensions}

  \author{K. J. Schrenk}
    \email{jschrenk@ethz.ch}
    \affiliation{Computational Physics for Engineering Materials, IfB, ETH Zurich, Schafmattstrasse 6, CH-8093 Zurich, Switzerland}

  \author{N. A. M. Ara\'ujo}
    \email{nuno@ethz.ch}
    \affiliation{Computational Physics for Engineering Materials, IfB, ETH Zurich, Schafmattstrasse 6, CH-8093 Zurich, Switzerland}

  \author{H. J. Herrmann}
	\email{hans@ifb.baug.ethz.ch}
	\affiliation{Computational Physics for Engineering Materials, IfB, ETH Zurich, Schafmattstrasse 6, CH-8093 Zurich, Switzerland}
	\affiliation{Departamento de F\'isica, Universidade Federal do Cear\'a, Campus do Pici, 60451-970 Fortaleza, Cear\'a, Brazil}

\begin{abstract}
  The Gaussian model of discontinuous percolation, recently introduced by Ara\'ujo and Herrmann [{\it Phys. Rev. Lett.}, {\bf 105}, 035701 (2010)], is numerically investigated in three dimensions, disclosing a discontinuous transition. 
  For the simple-cubic lattice, in the thermodynamic limit, we report a finite jump of the order parameter, $J=0.415 \pm 0.005$.
  The largest cluster at the threshold is compact, but its external perimeter is fractal with fractal dimension $d_A = 2.5 \pm 0.2$. 
  The study is extended to hypercubic lattices up to six dimensions and to the mean-field limit (infinite dimension).
  We find that, in all considered dimensions, the percolation transition is discontinuous.
  The value of the jump in the order parameter, the maximum of the second moment, and the percolation threshold are analyzed, revealing interesting features of the transition and corroborating its discontinuous nature in all considered dimensions.
  We also show that the fractal dimension of the external perimeter, for any dimension, is consistent with the one from bridge percolation and establish a lower bound for the percolation threshold of discontinuous models with finite number of clusters at the threshold.
\end{abstract}

\maketitle

\section{Introduction}

Percolation, one of the most famous models in statistical physics, has been extensively considered as a paradigm to study connectivity and transport \cite{Broadbent57,*Stauffer94,*Sahimi94}. 
Recently, Achlioptas, D'Souza, and Spencer \cite{Achlioptas09} have proposed a best-of-two product rule for bond selection characterized by a more pronounced transition than in the random case, being apparently discontinuous \cite{Friedman09,*Hooyberghs11}. 
This model has been analyzed on several different graphs \cite{Ziff09,*Ziff10,*Radicchi09,*Radicchi10,*Cho09,*Cho10,*Rozenfeld10,*Pan10,Fortunato11} and the ambiguous reported results raised controversy about the nature of the transition \cite{daCosta10,Tian10,Grassberger11,Nagler11,Lee11,Chen11,Riordan11}, with analytical \cite{daCosta10,Riordan11} and numerical \cite{daCosta10,Fortunato11} results showing the continuous nature of the transition in the original best-of-two product rule.
Several different models have been studied to shed light on the main mechanisms leading to a discontinuous percolation transition \cite{Manna10,Moreira10,DSouza10,*Christensen11,Araujo10}. 
A generalization to a best-of-$m$ product rule has also been proposed \cite{Andrade11b} and a tricritical point found when explosive percolation, obtained with $m=10$, is diluted with classical percolation on a square lattice \cite{Araujo11}.

Ara\'ujo and Herrmann \cite{Araujo10} introduced two models yielding clear discontinuous transitions: the largest cluster and the Gaussian models. 
The study of the former discloses the control of the largest cluster as a way to obtain homogenization of the cluster sizes and, consequently, an abrupt transition. 
Since the properties of the best-of-two product rule depend crucially on the topology \cite{Achlioptas09,Ziff09,Ziff10,Radicchi09,Radicchi10,Cho09}, in this work, we study the Gaussian model on hypercubic lattices up to dimension six and in the mean-field limit (infinite dimension).
We report that, for all dimensions, the Gaussian rule leads to a discontinuous transition at the percolation threshold and that the fractal dimension of the largest-cluster external perimeter is compatible with the one reported for bridge percolation \cite{Araujo11b}.

This manuscript is organized in the following way. 
In the next section we describe the Gaussian model and analyze its properties on the simple-cubic lattice.
In Sec.\,\ref{sec::hd} the study of the model is extended to higher dimensions. 
We leave the final remarks for Sec.\,\ref{sec::fin}.

\section{The Gaussian model on the simple-cubic lattice}

We start by considering a simple-cubic lattice with linear size $L$ and periodic boundary conditions in all directions. In the initial configuration, all the $3N$ bonds are empty, such that there are $N=L^3$ clusters of size unity. At each iteration, a new bond is randomly chosen among the empty bonds and occupied with probability
\begin{equation}\label{eqn::pgm}
  \min\left\{1,\exp \left[ -\alpha\left( \frac{s-\bar s}{\bar s} \right)^2 \right]\right\} \  \  ,
\end{equation}
where $s$ is the size of the cluster that would be formed by occupying the selected bond and $\bar s$ is the average number of sites per cluster if the bond would be occupied. For bonds which connect sites belonging to the same cluster, $s$ is taken as twice the size of the cluster. $\alpha$ is a parameter of the model which, for the sake of simplicity, we take equal to unity. The proposed method promotes the homogenization of the cluster sizes by suppressing the formation of clusters differing significantly, in size, from the average.

\begin{figure}
	\begin{tabular}{cc}
	\includegraphics[width=0.499\columnwidth]{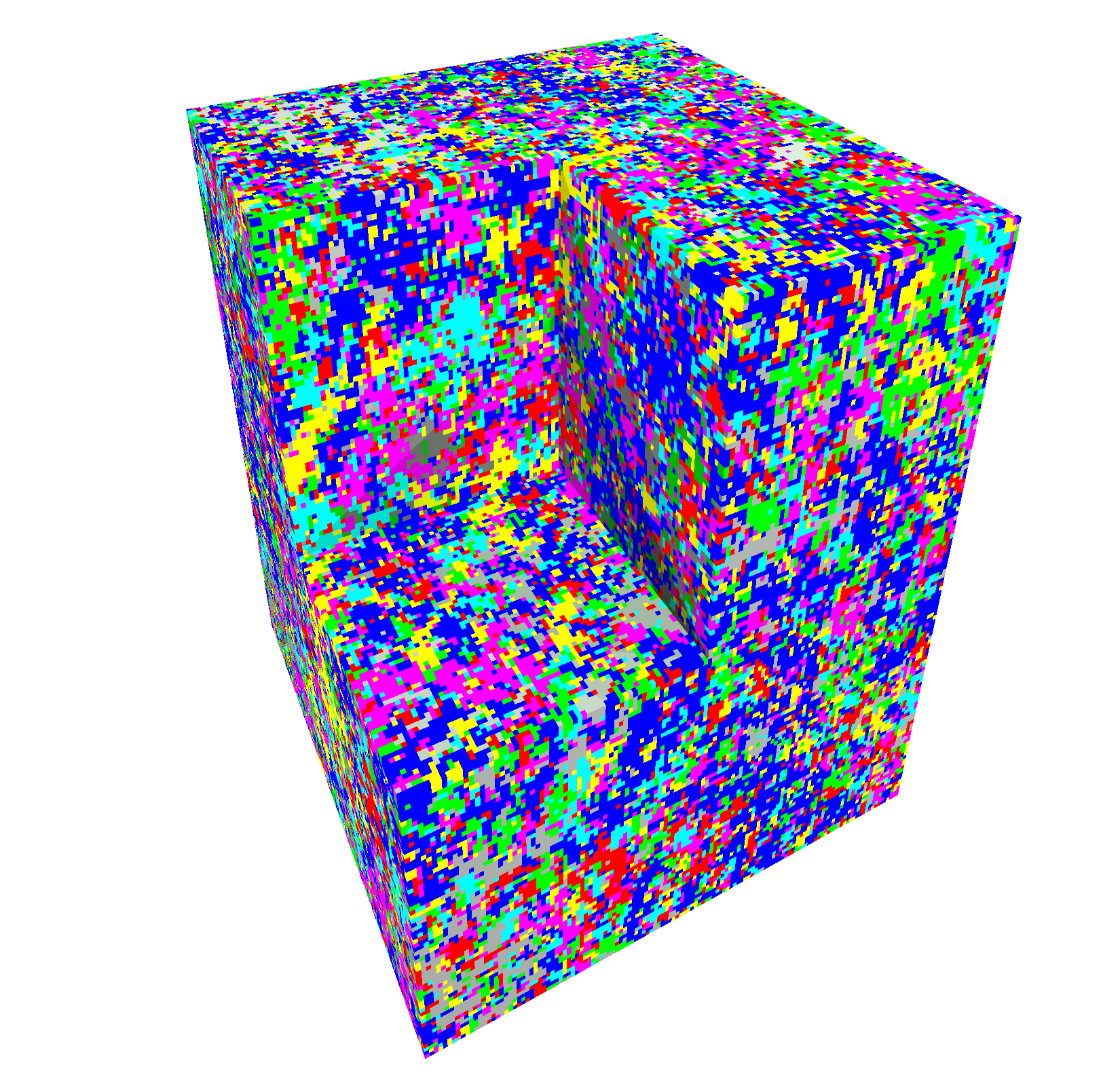}&
	\includegraphics[width=0.499\columnwidth]{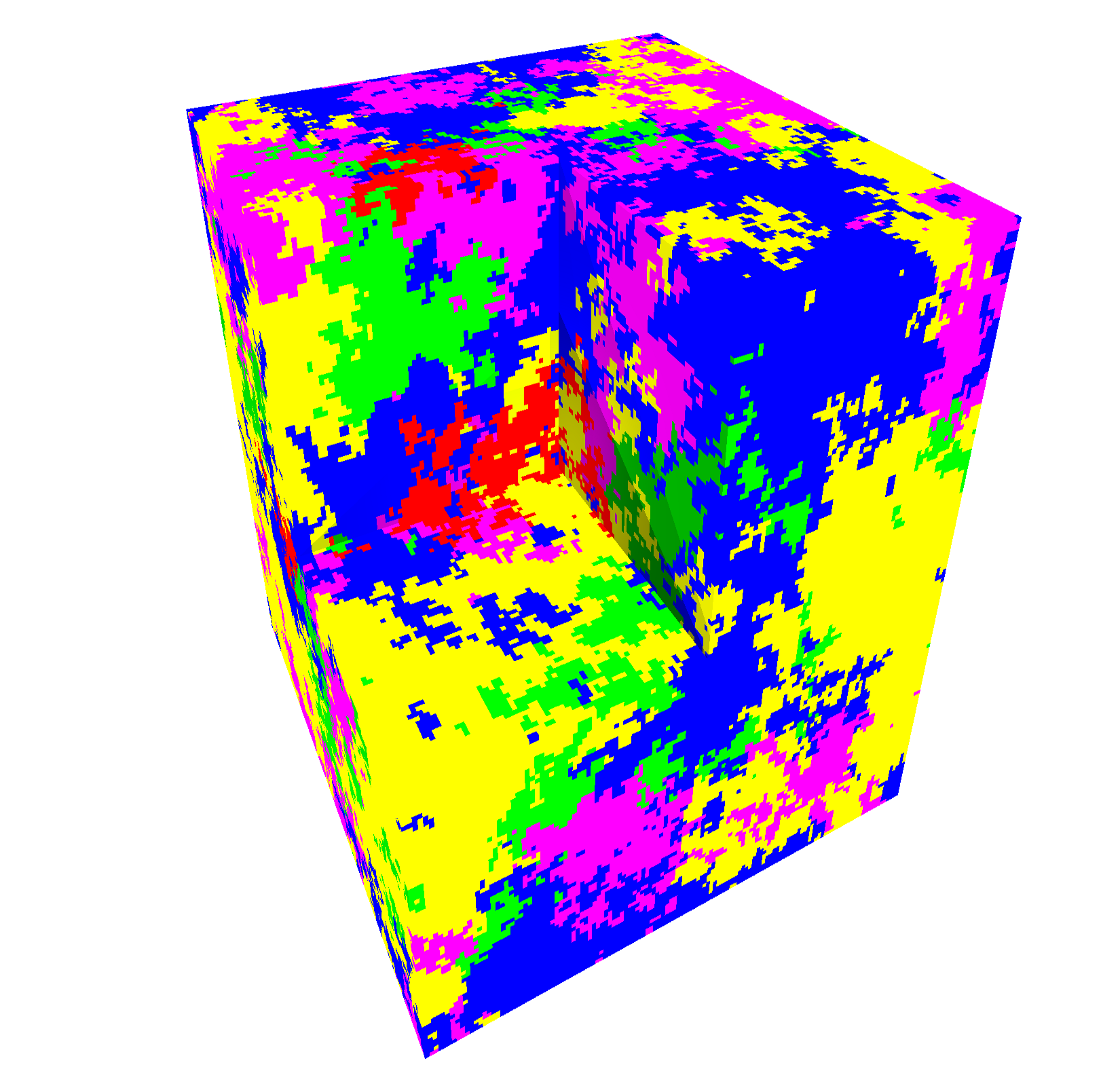}\\
	(a)&(b)\\
	\includegraphics[width=0.499\columnwidth]{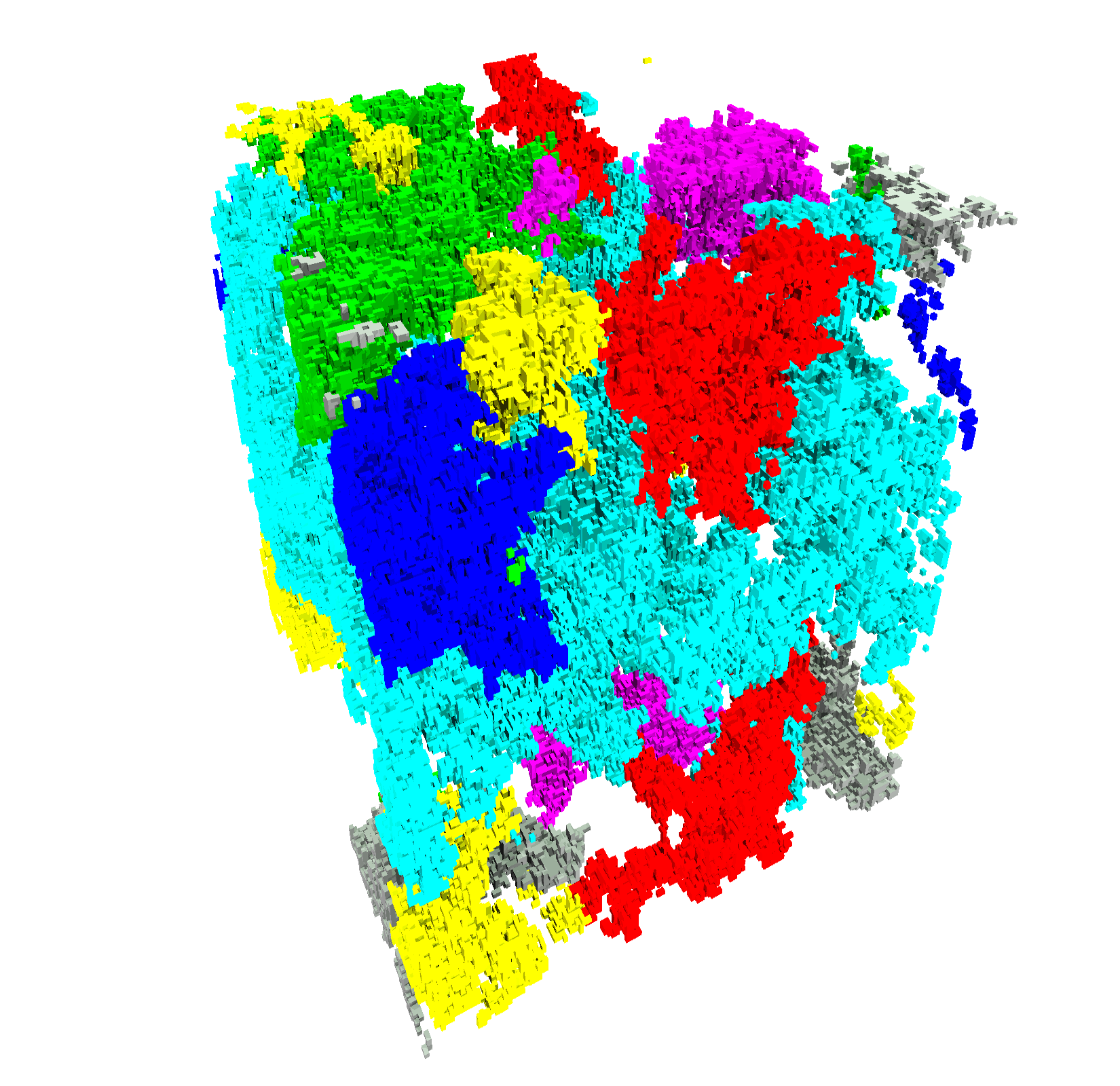}&
	\includegraphics[width=0.499\columnwidth]{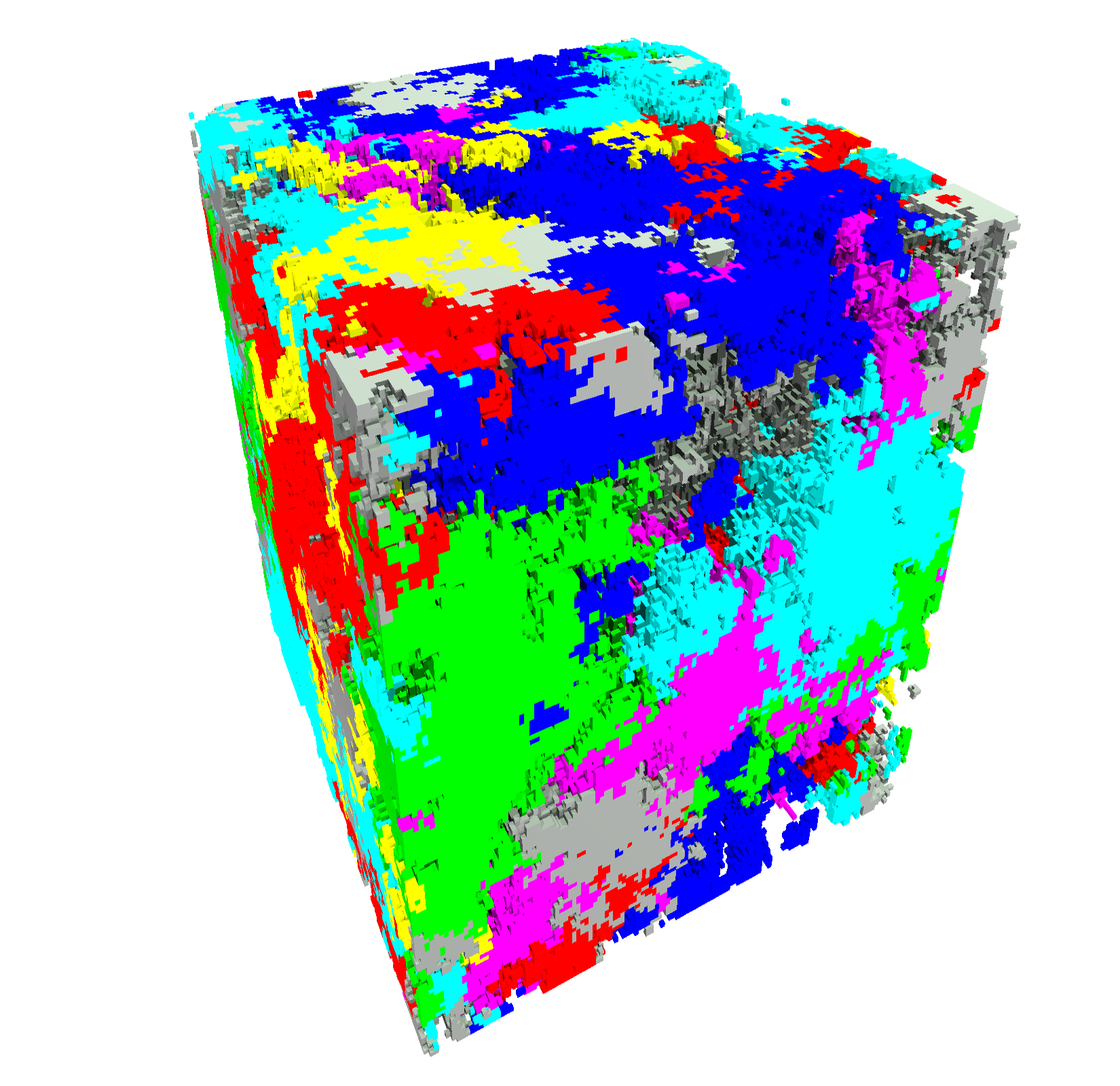}\\
	(c)&(d)
	\end{tabular}
	\caption{
	(color online) Snapshots of (a) classical percolation and (b) the Gaussian model of discontinuous percolation, at the percolation threshold, on a simple-cubic lattice with $128^3$ sites. To enhance visibility, the front cube of size $64^3$ has been left out in both pictures. While in the classical case clusters have very different size and a fractal shape, for the Gaussian model clusters are rather compact and with a characteristic size. The seven largest clusters of the configurations are shown in (c) for classical percolation and in (d) for the Gaussian model.
	\label{fig::snap}
	}
\end{figure}

The difference between classical percolation and the Gaussian model can be seen qualitatively in Fig.\,\ref{fig::snap}, where we show snapshots for both models, on the simple-cubic lattice, at the respective percolation thresholds. For classical percolation, Fig.\,\ref{fig::snap}(a) and (c), the clusters are fractal and of very different sizes, following a power-law distribution \cite{Stauffer94}, whereas for the Gaussian model (Fig.\,\ref{fig::snap}(b) and (d)) clusters are rather compact and of comparable size. Note that, while for classical percolation, the number of clusters is large, $\approx 0.27N$, the number of clusters in the Gaussian model is significantly smaller. 
In fact, as we show here, in the thermodynamic limit, the Gaussian model is characterized by a finite number of macroscopic clusters at the threshold.

\begin{figure}
  \includegraphics[width=\columnwidth]{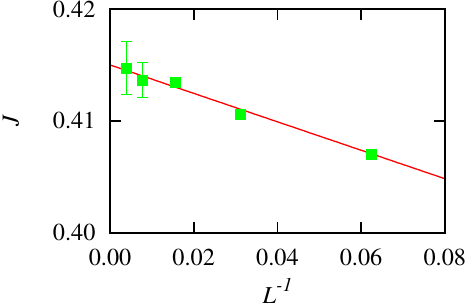}
  \caption{
    (color online) System size dependence of the maximum jump of the order parameter, $J$ ($\blacksquare$), for the Gaussian model of discontinuous percolation on the simple-cubic lattice. 
    In the classical case, the change in the order parameter shrinks with the system size and is zero in the thermodynamic limit (not shown), whereas in the discontinuous case, a finite, non-zero value ($J=0.415 \pm 0.005$) is obtained in this limit. Note that, whereas for the product rule and classical percolation the size of the jump decreases with the system size, for the Gaussian model it even slightly increases for the same range of system sizes.
    Results have been averaged over $10^6$ samples for the smallest system size and $1.1\times 10^3$ samples for the largest one. 
    Random numbers have been generated with the algorithm proposed in Ref.\,\cite{Ziff97}. 
    To identify the clusters and keep track of their properties we have considered the labeling scheme proposed by Newman and Ziff \cite{Newman00,*Newman01}, related to the Hoshen--Kopelman algorithm \cite{Hoshen76}.
    \label{fig::MaxJump}
  }
\end{figure}

To analyze the behavior of the order parameter, namely the fraction of sites in the largest cluster, we measure, for each sample, its jump $J$, defined as the maximum change obtained as one sequentially occupies bonds in the system \cite{Manna10,Nagler11}. 
For every considered linear system size $L$, we average the jump and the fraction of occupied bonds $p$, at which it occurs $p_{c,J}$ over several configurations.
We take the latter as an estimator for the threshold in the thermodynamic limit.
Recently, Lee \textit{et al.} \cite{Lee11} defined it as the upper pseudo-transition point and used it to pin down the threshold.
Plotting $J$ as a function of $L^{-1}$ reveals that for the Gaussian model, in the thermodynamic limit, the jump has a finite value of $J=0.415 \pm 0.005$ (see Fig.\,\ref{fig::MaxJump}), as expected for a discontinuous transition. This result is in contrast to the ones for classical percolation and the product rule where, for the same range of system sizes, the size of the jump diminishes and eventually vanishes in the thermodynamic limit \cite{Nagler11}.

\begin{figure}
  \includegraphics[width=\columnwidth]{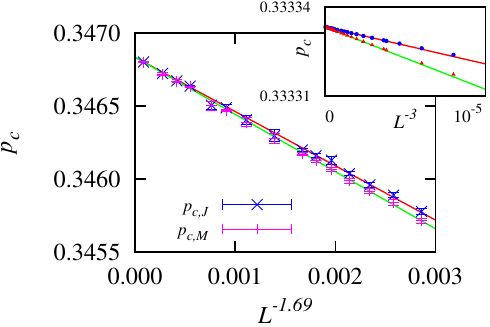}
  \caption{
    (color online) Threshold $p_c$, as a function of the inverse linear system size $L^{-a}$, for the Gaussian model of discontinuous percolation on the simple-cubic lattice. 
    $p_{c,J}$ $(\times)$ stands for the average fraction of occupied bonds at which the jump occurs and $p_{c,M}$ $(+)$ for the position of the maximum of the second moment of the cluster size distribution without the contribution of the largest cluster.
    The percolation threshold is estimated to be $0.3468 \pm 0.0005$. Results have been averaged over $10^3$ samples for the smallest system size ($32^3$ sites) and $10^2$ samples for the largest one ($256^3$ sites).
    The inset shows the same for the case where only merging bonds are considered; this is, all clusters are trees (loopless). In this case, the threshold estimators [$p_{c,J}$ $(\bullet)$, $p_{c,M}$ $(\blacktriangle)$] depend asymptotically linearly on $L^{-3}$. The threshold for the loopless case is estimated to be $0.3333 \pm0.0004$. Results have been averages over $10^3$ samples for the smallest system site ($50^3$ sites) and $10^2$ samples for the largest one ($512^3$ sites). Error bars are smaller than the symbol size.
    \label{fig::pc_est}
  }
\end{figure}

To determine the threshold $p_c$, two different estimators have been considered: the average fraction of occupied bonds at which the jump occurs $p_{c,J}$ \cite{Manna10,Nagler11}, and the position $p_{c,M}$ \cite{Ziff02}, of the maximum in the second moment of the cluster size distribution, excluding the contribution of the largest cluster (of size $s_\text{max}$), 
\begin{equation}
M_2' = M_2 - s_\text{max}^2/N \  \  ,
\end{equation}
where $M_2 = \sum_i s_i^2/N$ and $s_i$ is the size of cluster $i$. Figure \ref{fig::pc_est} shows the system size dependence of both estimators on the simple-cubic lattice. Asymptotically, a dependence on $L^{-a}$ is found, compatible with using the same exponent $a=1.69 \pm 0.10$ for $p_{c,J}$ and $p_{c,M}$. The estimators are extrapolated to the thermodynamic limit, given by $L^{-a}\to 0$, and combining both methods yields $p_c = 0.3468 \pm 0.0005$. Note that, for an equilibrium first-order transition, $a=d$ \cite{Binder81,*Binder84}. To shed light on the obtained value, we investigate the dependence of the threshold estimators on the system size, under the constraint that only merging bonds are considered; this is, all clusters are trees (loopless). In the inset of Fig.\,\ref{fig::pc_est} we see $p_{c,J}$ and $p_{c,M}$ as a function of $L^{-3}$. One observes that in this case, $a=d$. Therefore, in the Gaussian model $a$ differs from $d$ due to internal bonds which do not influence the cluster structure or the size of the jump.

\begin{figure}
  \includegraphics[width=\columnwidth]{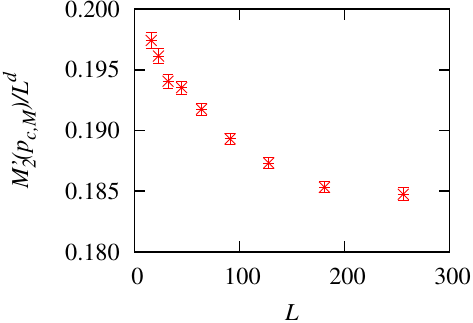}
  \caption{
	(color online) Maximum of the second moment of the cluster size distribution per lattice site, $M_2'(p_{c,M})/L^d$ ($\ast$), as a function of the linear system size $L$. The second moment per lattice site tends toward a constant value, as expected for a discontinuous transition. Results have been averaged over $10^3$ samples for the smallest system size and $2.4\times10^2$ samples for the largest one.
  \label{fig::SecondMoment}
	}
\end{figure}

In Fig.\,\ref{fig::SecondMoment} we see the size dependence of the maximum of the second moment per lattice site. For every sample, we measure the maximum $M_2'(p_{c,M})/L^d$ and average over all samples.  For large system sizes, this quantity is constant, as expected for a discontinuous transition.

\begin{figure}
  \includegraphics[width=\columnwidth]{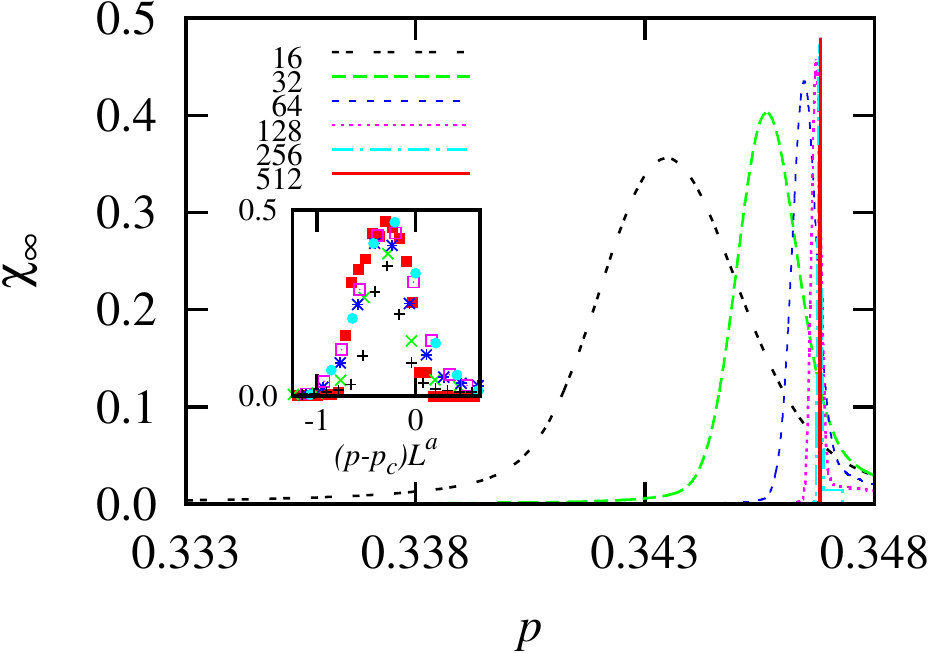}
  \caption{
  (color online) Standard deviation of the order parameter $\chi_\infty$ as a function of the bond occupation fraction $p$ for different linear system sizes $L$. One observes that the peak increases and narrows with the system size. In the inset we see $\chi_\infty$ as a function of the scaling variable $(p-p_c)L^a$, with $a=1.64$, for different linear system sizes $L$: $16$ ($+$), $32$ ($\times$), $64$ ($\ast$), $128$ ($\square$), $256$ ($\bullet$), and $512$ ($\blacksquare$). Results have been averaged over $10^{8}$ samples for the smallest system size and $28$ samples for the largest one.
  \label{fig::std_sc}
  }
\end{figure}

The scaling behavior of the standard deviation of the order parameter, defined as
\begin{equation}
  \chi_\infty = \sqrt{\langle s_\text{max}^2 \rangle - \langle s_\text{max} \rangle^2} \big/ N  \  \ ,
\end{equation}
is shown in Fig.\,\ref{fig::std_sc}. For large systems, the maximum of
$\chi_\infty$ tends toward a constant value. These results are a
strong evidence of a discontinuous transition since a nonzero value of
$\chi_\infty$ is obtained at the transition point, as expected in the
presence of a jump in the order parameter
\cite{Binder81,*Binder84,Odor04}. In addition, the plot is consistent with the exponent $a = 1.69\pm 0.10$ and $p_c = 0.3468\pm 0.0005$, as determined from the finite size scaling of $p_{c,J}$ and $p_{c,M}$.

We measure at $p_c$ the external perimeter of the largest cluster $A$. The external perimeter is defined as the number of sites which do not belong to the largest cluster but are nearest neighbors of sites in this cluster \cite{Voss84}.
One observes that, at the threshold, the external perimeter of the largest cluster scales asymptotically with the system size as $A\sim L^{d_A}$, where $d_A = 2.5 \pm 0.2$ (see also Fig.\,\ref{fig::ghd_surf}).
On the square lattice, the fractal dimension of the external perimeter was shown to be related to several other models \cite{Araujo10,Andrade11}.
The value reported here for the simple-cubic lattice agrees within its error bars with the one for watersheds and the optimal path cracking \cite{Cieplak94,*Cieplak96,*Andrade09,*Oliveira11,*Fehr09,*Fehr11} as well as with the set of bridges in bridge percolation \cite{Araujo11b}.
Clusters at the threshold are compact with fractal external perimeter, as was also reported for $2D$ \cite{Araujo10} and for irreversible aggregation at high concentration \cite{Kolb87}.

\section{Higher dimensions and mean-field behavior\label{sec::hd}}

The Gaussian model yields a discontinuous percolation transition in two and three dimensions. How does the nature of the transition depend on the dimensionality of the system? To address this question, we consider the Gaussian model on hypercubic lattices up to $d=6$, the upper critical dimension of classical percolation \cite{Toulouse74,*Chayes87}. In addition, the mean-field behavior of the Gaussian model is investigated. In the latter case, we take a system with $N$ sites which can be fully interconnected giving a total of $N(N-1)/2$ links, and we add links between sites with probability given by Eq.\,(\ref{eqn::pgm}). For this system, $p$ is defined as the average number of links per site. Occupying links randomly, without any additional rule, would recover Erd\H{o}s--R\'enyi percolation, where $p_c = 1/2$ (see, for example, Ref.\,\cite{Albert02,*Dorogovtsev02}).

\begin{figure}
	\includegraphics[width=\columnwidth]{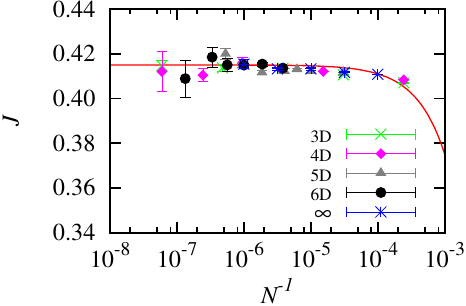}
	\caption{
	(color online) System size dependence of the jump $J$, for the Gaussian model of discontinuous percolation on the hypercubic lattice of dimension $3$ ($\times$), $4$ ($\blacklozenge$), $5$ ($\blacktriangle$), and $6$ ($\bullet$), as well as in the mean-field case ($\ast$). In the limit $N^{-1}\to 0$, the jump has within the error bars the same finite value $0.415$ for all considered graphs. The solid line is a guide to the eye and of the form $0.415-40N^{-1}$. For the sake of comparison, we plot the jump as a function of the inverse system size $N^{-1}$. Results have been averaged over $10^7$ samples for the smallest system size and at least $10$ samples for the largest one.
	\label{fig::GaussHDMaxJump}
	}
\end{figure}

Figure \ref{fig::GaussHDMaxJump} shows the jump $J$, as a function of the inverse system size $N^{-1}$, for $3$ to $6$ dimensions and for mean-field. 
We observe that, in the thermodynamic limit, $J$ has, within the error bars, the same finite value in all considered dimensions, consistent with the value found in three dimensions, $J=0.415\pm 0.005$ (see Fig.\,\ref{fig::MaxJump}).
In general, we expect for a discontinuous percolation transition to find few macroscopic clusters at the threshold, as initially discussed by Friedman and Landsberg \cite{Friedman09,*Hooyberghs11}. 
Nagler, Levina, and Timme \cite{Nagler11} have added that, for strongly discontinuous transitions, where the largest cluster cannot grow directly, the number of clusters is finite and the transition occurs when the two largest clusters merge.
The jump is then bounded by two limits: either the clusters have the same size, giving $J=1/2$, which is the largest possible jump size in discontinuous percolation, or the largest cluster is of size $\approx 2/3$ and the second largest of size $\approx 1/3$, giving $J=1/3$. 
The latter case corresponds to situations where the second cluster merges with the third one and becomes the largest one, of size $2/3$, called overtaking in Ref.\,\cite{Nagler11}, merging later with the one of $1/3$. 
On the other hand, the former results from four clusters of equal size which merge in pairs. 
This case is expected for the global competition proposed in Ref.\,\cite{Manna10}, in the mean-field limit. 
The Gaussian model at any dimension also promotes the homogenization of the cluster sizes and the values of the jump are within the proposed interval.
The same idea can be considered to understand the behavior of the maximum of $M_2'/N$, taking place at $p_{c,M}$ which is our second estimator.

\begin{figure}
	\includegraphics[width=\columnwidth]{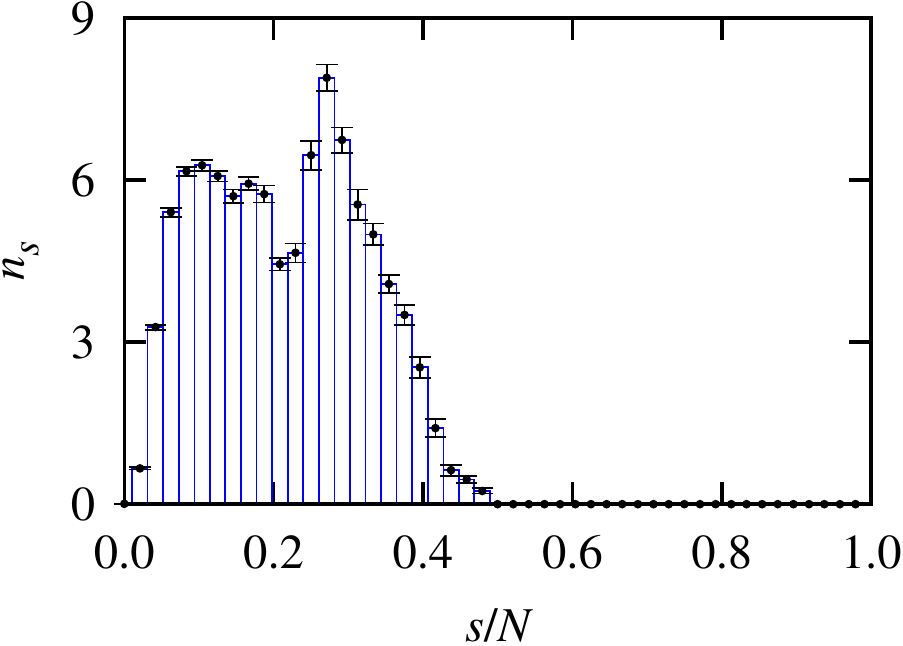}
	\caption{
	(color online) Cluster-size distribution for the Gaussian model on the simple-cubic lattice. The fraction $n_s$ ($\bullet$) of clusters of size $s$ times the size of the cluster $s/N$ is shown as a function of $s/N$. One observes that the distribution is bimodal, as expected for discontinuous transitions. The system size is $64^3$ sites, results have been averaged over $1.6\times 10^5$ samples, and error bars are indicated.
	\label{fig::GaussClusterSizeDistribution3D}
	}
\end{figure}

In Fig.\,\ref{fig::GaussClusterSizeDistribution3D} we see the cluster-size distribution for the Gaussian model on the simple-cubic lattice at the percolation threshold, $p=p_c$.
As previously observed in $2D$ \cite{Araujo10}, a bimodal distribution is obtained, in contrast with the power-law behavior observed for random percolation \cite{Stauffer94} and the best-of-two product rule \cite{Ziff10}. Since the contribution of the largest cluster is neglected, there is a cut-off at $s/N=0.5$.

As in three dimensions, we also determine the percolation threshold for the Gaussian model in dimensions $4$, $5$, and $6$, as well as in the mean-field limit, by combining both estimators: $p_{c,J}$ and $p_{c,M}$. 
Table \ref{tab::pcghd} shows the threshold values $p_c$, for different dimensions. 
One observes that $p_c$ decreases with the dimension, though it remains always above the values for classical percolation \cite{Vyssotsky61,*Sur76,*Grassberger86,*Grassberger92,*Grassberger03,*Gouyet88,*Lorenz98,*Paul01,*Dammer04} (shown in the same table for comparison). 
For the Gaussian model in the mean-field limit we find $p_c$ to be compatible with unity but note that, in this case, $p$ is defined as the fraction of bonds per site and not the fraction of occupied bonds, as in the lattice case.
Below we establish a lower bound for the $p_c$ of models yielding a discontinuous percolation transition with finite number of clusters at the threshold.

Consider an arbitrary percolation model which starts with isolated clusters of unit size, adding bonds sequentially until a certain fraction of occupied bonds is reached. Let us denote by $c(p)$ the number of clusters at a given fraction of occupied bonds $p$.
At each iteration, added bonds to the system can be merging bonds -- connecting two clusters -- or redundant bonds -- connecting nodes of the same cluster \cite{Moreira10}.
Only the former bonds change $c(p)$.
The number of clusters reduces by one if the bond is a merging bond and does not change if it is a redundant bond. If we define $r(p)$ as the probability that an added bond is redundant, then
\begin{equation}
  \frac{\mathrm{d}c}{\mathrm{d}b} = -\left[1-r(p)\right] \  \  ,
\end{equation}
where $b=pNd$ is the number of occupied bonds in a $d$-dimensional hypercubic lattice. 
We now take the limit $N\to \infty$,
\begin{equation}
  \lim_{N\to\infty} \frac{\mathrm{d}c}{\mathrm{d}pN} = -d\left[1-r(p)\right] \  \ .
\end{equation}
Integrating over the interval $0\leq p \leq p_c$ gives
\begin{equation}
  \lim_{N\to \infty} \left[c(p_c)-c(0)\right]/N = -d \int_0^{p_c}\mathrm{d}p\left[1-r(p)\right] \  \  .
\end{equation}
This equation is valid for any percolation model regardless the nature
of the transition. For example, the percolation threshold for the
classical tree-like case can be obtained by taking $r(p)\equiv0$ and the
proper number of clusters at the threshold \cite{Ziff97b,Temperley71,Baxter78}.

Assuming the cases where the homogenization of the cluster sizes leads to a finite number of clusters at $p_c$ \cite{Nagler11}, and since $c(0) = N$,
\begin{equation}
  p_c = 1/d +  \int_0^{p_c}\mathrm{d}p\, r(p) \  \  ,
\end{equation}
we obtain that $p_c \geq 1/d$. 
Note that for tree-like models $r(p)\equiv0$ and, if $c(p_c)/N \to 0$ as $N\to\infty$, $p_c = 1/d$. 
For the Hamiltonian model of explosive percolation, introduced by Moreira \emph{et al.}\ \cite{Moreira10}, the same result was derived in the mean-field limit in an independent way and numerically observed in the lattice.
The value reported by Manna and Chatterjee \cite{Manna10} for the case with global competition is also consistent with this result. 
Both the lower bound for the threshold and the solution $c(p<p_c)/N\approx 1-pd$, obtained for vanishing small probability of redundant bonds, are consistent with the numerical results for the Gaussian model.
For increasing dimension the contribution of redundant bonds decreases and $p_c$ approaches $1/d$ (compare Tab.\,\ref{tab::pcghd}). 
In the mean-field limit, this asymptotic behavior also agrees within error bars with the reported results.
The transition is obtained when the number of added bonds equals the number of nodes $N$.
Since the maximum number of bonds is $N(N-1)/2$, the fraction is zero in the thermodynamic limit.
\begin{table}
  \caption{
    Percolation threshold $p_c$ for the Gaussian model of discontinuous percolation on the hypercubic lattice of dimension $d$ and in the mean-field limit. For comparison, the percolation thresholds for classical percolation are shown in the third column. Note that, in this table, for all models, $p$ is defined as the fraction of occupied bonds, $p=t/(Nd)$ for hypercubic lattices and $p=t/N$ for mean field, where $t$ is the number of occupied bonds in the system.
    \label{tab::pcghd}
    }
    \begin{tabular}{llll}
      \hline\hline
      $d$ & $p_c$ & $d_A$ & $p_c$ classic \\
      \hline
      $2$ & $0.56244(6)$ \cite{Araujo10} & $1.23(3)$ \cite{Araujo10} & $1/2$ \cite{Stauffer94} \\
      $3$ & $0.3468(5)$ & $2.5(2)$ & $0.2488126(5)$ \cite{Lorenz98} \\
      $4$ & $0.254(2)$ & $3.6(4)$ & $0.1601314(13)$ \cite{Grassberger03} \\
      $5$ & $0.202(2)$ & $4.9(7)$ & $0.118172(1)$ \cite{Grassberger03} \\
      $6$ & $0.168(3)$ & $5.9(8)$ & $0.0942019(6)$ \cite{Grassberger03} \\
      $\infty$ & $1.000(2)$ & & $1/2$ \cite{Albert02,*Dorogovtsev02} \\
      \hline
    \end{tabular}
\end{table}
\begin{figure}
  \includegraphics[width=\columnwidth]{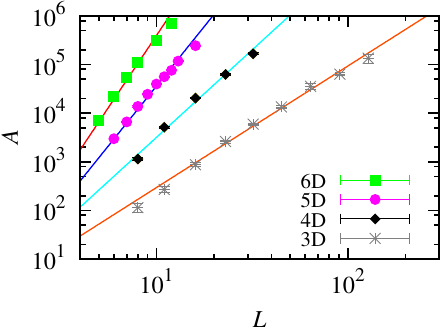}
  \caption{
    (color online) System size dependence of the external perimeter of the largest cluster $A$, at the threshold, for the Gaussian model of discontinuous percolation on the hypercubic lattice of dimension $3$ ($\ast$), $4$ ($\blacklozenge$), $5$ ($\bullet$), and $6$ ($\blacksquare$).
    One observes that $A$ asymptotically scales with the system size as $A\sim L^{d_A}$. 
    The solid lines have slopes of $5.9 \pm 0.8$, $4.9 \pm 0.7$, $3.6 \pm 0.4$, and $2.5 \pm 0.2$, respectively. 
    Results have been averaged over $10^3$ samples.
    \label{fig::ghd_surf}
  }
\end{figure}

Measuring the size dependence of the largest-cluster external perimeter $A$, at the percolation threshold $p_c$, its fractal dimension is obtained for dimensions $4$, $5$, and $6$ (see Fig.\,\ref{fig::ghd_surf}). 
For increasing dimension, $d_A$ seems to approach $d$ (see Tab.\,\ref{tab::pcghd}). 
These exponents are compatible with the ones found for bridge percolation, corroborating their equivalence \cite{Araujo11b}.

\section{Final remarks\label{sec::fin}}

In summary, in this work we studied the Gaussian model of discontinuous
percolation in three and higher dimensions.  We disclose that, for any
considered dimension, the percolation transition is abrupt and
characterized by a discontinuity in the order parameter, which within
error bars is independent on dimension.  We identify the homogenization
of cluster sizes and favoring of merging bonds as the key mechanisms
leading to such an abrupt transition
\cite{Araujo10,Moreira10,Manna10,Nagler11}.  For discontinuous
percolation models with a finite number of macroscopic clusters at the
threshold, we establish a lower bound for $p_c$ as well as a relation
between $p_c$ and the probability of selecting a redundant bond.
Studying different dimensions we show that clusters are compact with a
fractal perimeter with the same dimension as bridge percolation
\cite{Araujo11b}, which is also related to watersheds and the optimal
path cracking \cite{Fehr09,*Andrade09,*Oliveira11}.  Although all
numerical indications point in that direction we have no formal proof
whether the upper-critical dimension for the Gaussian model is six, like
in the classical case. In addition, the meaning of the non-trivial
finite-size scaling exponent $a=1.69\pm 0.10$, consistent for both
estimators of $p_c$, is still puzzling;  an analytical treatment of this
exponent would be interesting.  Studies of this model have taken
$\alpha=1$, in Eq.\,(\ref{eqn::pgm}), but it would be interesting to
investigate other cases since for $\alpha=0$ the model boils down to the
classical percolation model. It would be interesting to study how the
described properties depend on $\alpha$.  Future work might also consist
in studying the behavior of other models of explosive percolation, like
the global competition proposed by Manna and Chatterjee \cite{Manna10}
and the BFW model discussed by Chen and D'Souza \cite{Chen11}, in
different dimensions.
Besides, all known models with a
discontinuous transition imply global information. It is still an open
question if a discontinuous percolation transition can be obtained with
only local rules.

\begin{acknowledgments}
We thank Robert Ziff for his useful comments. We acknowledge financial support from the ETH Competence Center Coping with Crises in Complex Socio-Economic Systems (CCSS) through ETH Research Grant CH1-01-08-2.
We also acknowledge the Brazilian agencies CNPq, CAPES and FUNCAP, and the Pronex grant CNPq/FUNCAP, for financial support.
\end{acknowledgments}

\bibliography{ghd}

\end{document}